\documentclass[12pt]{article}

\newcommand{\citep}{\cite}
\usepackage{amsfonts}
\usepackage{amssymb}
\usepackage{epsfig}
\usepackage{rotating}
\usepackage{lineno}
\usepackage{color}
\usepackage{amsmath}
\usepackage{mathrsfs}
\usepackage{hyperref}
\usepackage{setspace}
\usepackage{grffile}
\usepackage{multirow}
\usepackage{verbatim}   
\usepackage{url}

\makeatletter
\def\url@leostyle{%
  \@ifundefined{selectfont}{\def\UrlFont{\sf}}{\def\UrlFont{\small\ttfamily}}}
\makeatother
\usepackage{hyperref}
\hypersetup{pagebackref, bookmarks=true,
citecolor=blue, urlcolor=blue}
\let\epsilon=\varepsilon

\newcommand{\Ro}{\ensuremath{{\mathcal R}_0\;}}

\newcommand{\eg}{\emph{e.g.,} }

\begin{document}


\title{Global Spatio-temporal Patterns of Influenza in the Post-pandemic Era}

\author{
 Daihai He$^1$, Roger Lui$^{2}$,  Lin Wang$^{3}$, Chi Kong Tse$^4$, Lin Yang$^{5}$ and Lewi Stone$^{6\star}$
\vspace{0.2cm}\\
{\footnotesize $^1$ Department of Applied Mathematics, Hong Kong Polytechnic University, Hong Kong (SAR) China}\\
{\footnotesize $^2$ Department of Mathematical Sciences, Worcester Polytechnic Institute, 100 Institute Road}\\
{\footnotesize  Worcester, MA 01609, United States}\\
{\footnotesize $^3$ School of Public Health, Li Ka Shing Faculty of Medicine, University of Hong Kong,                                                         Hong Kong (SAR) China}\\
{\footnotesize $^4$ Department of Electronic and Information Engineering, Hong Kong Polytechnic University}\\
{\footnotesize Hong Kong (SAR) China}\\
{\footnotesize $^5$ School of Nursing, Hong Kong Polytechnic University, Hong Kong (SAR) China}\\
{\footnotesize $^6$ School of Mathematical and Geospatial Sciences, RMIT University, Melbourne, 3000, Australia}\\
{\footnotesize $\star$ lewistone2@gmail.com}
}

\maketitle

\section*{Abstract}
We study the global spatio-temporal patterns of influenza dynamics. This is achieved  by analysing and modelling   weekly laboratory confirmed cases of influenza A and B from 138 countries between January 2006 and May 2014. The data were obtained from FluNet,  the surveillance network compiled by the the World Health Organization.  We report a pattern of {\it skip-and-resurgence} behavior between the years 2011 and 2013 for influenza H1N1pdm, the strain responsible for the 2009 pandemic, in Europe and Eastern Asia. In particular, the expected H1N1pdm epidemic outbreak in 2011 failed to occur (or ``skipped'') in many countries across the globe, although an outbreak occurred in the following year.  We also report a pattern of {\it well-synchronized} 2010 winter wave of H1N1pdm in the Northern Hemisphere countries, and a pattern of replacement of strain H1N1pre by H1N1pdm between the 2009  and 2012 influenza seasons. Using both a statistical and a mechanistic mathematical model, and through fitting the data of 108 countries (108 countries in a statistical model and 10 large populations with a mechanistic model), we discuss the mechanisms that are likely to generate these events taking into account the role of multi-strain dynamics. A basic understanding of these patterns has important public health implications and scientific significance.

\section{Introduction}\label{S:intro}

Seasonal influenza in temperate zones of the world are characterized by regular annual epidemics for most of the last fifty years\citep{CoxSubb00,Pott01,Simo99}.  According to historical reports, however, this  annual periodicity was less apparent in the past.  Between 1855 and 1889, influenza was not widely experienced and believed to have caused few deaths in Britain \cite{Andr53}.  In the first half of the twentieth century,  seasonal influenza  seemed ``erratic as regards its occurrence in both time and space" \citep{Brow23}.  Between 1920 and 1944 there were 16 widespread influenza (both A and B) epidemics in the United States, the remaining  eight years presumably being complete skips \citep{Commission}. In the same period in the United States  ``visitations of influenza B ...  tended to come every four to six years and those of A every two to three." \citep{Brow23,Andr53}.
 Similarly, while in recent years annual outbreaks are the norm,  skips by different  influenza subtypes (such as A and B) may unexpectedly occur, sometimes with one subtype temporarily replacing the other.  To add another layer of complexity,  the regular seasonal dynamic experience in the last decades can be grossly punctuated when a new  pandemic virus strain appears, as was the case  in 2009.  Understanding those factors that enhance annual dynamics, and those factors which break it up is a research direction that deserves more attention.  Even basic concepts concerning the competition between strains,  cross-immunity, the influence of climatic factors or the effects of a country's vaccination policy on the seasonal dynamics in the large, are poorly understood to date (\eg \cite{Tame+11}).

To help explore these sorts of complexities, in this paper, we are interested in characterizing the spatio-temporal dynamics of influenza  as they  occurred globally following the last 2009 pandemic.
A generic pattern easily identified for many countries in Europe is shown in Fig.~\ref{Fig:region} e,f, and g (red). There we see the initiation of the new H1N1pdm09  (referred to henceforth as H1N1pdm) pandemic in March 2009, followed by another major outbreak in September 2010.  Unusually  an H1N1pdm outbreak failed to appear at all in the ``skip year" of  2011/12, given that the strain was very new, although the outbreak returned and resurged in 2012/13.
As we will discuss shortly, this same pattern was generic to  many countries across Europe,  with slight differences from country to country.   A visualisation  of the extraordinary skip-year across 45 countries is given in Fig.~\ref{Fig:skip}.

Many of the features of the time series in Fig.~\ref{Fig:region} can be explained in terms  of basic epidemiological theory.
Briefly, when the new 2009 pandemic influenza strain  confronted a large susceptible human population it was able to generate a large-scale global epidemic. This placed in motion a succession of epidemic ``waves" that followed  one after the other.   Since infected individuals who recover from the disease gain temporary immunity,  each new epidemic wave also served to build up further the level of immunity  in the population. In effect, this served to reduce the number of susceptible individuals available for future infection.  At some point, when the number of available susceptibles fell below a threshold level, it became impossible for a new outbreak of the pandemic strain to trigger. This explains the ``skip" year in 2011/12 in which the strain  was mostly absent (see Fig.~\ref{Fig:skip}). The H1N1pdm strain resurged in 2012/13 presumably because  recovered individuals gradually lost their immunity, providing enough new susceptibles to trigger further outbreaks.
A systematic theory for understanding epidemic oscillations and skips has been developed over the last decade \citep{Ston+07}, which we will use to explain these long term dynamics.

Our detailed spatio-temporal analysis is based on time series obtained from FluNet, a comprehensive global surveillance tool for influenza developed by the World Health Organization (WHO) in 1997 \cite{Flah+98}, in which virological data are documented in real-time and publicly available.  When discussing and presenting the FluNet data it is convenient to use the following notation.  We denote H1N1pre as the pre-2009-pandemic seasonal A strains, H1N1pdm as the pandemic strain (H1N1pdm09) responsible for the 2009 influenza pandemic, and H3N2 as the seasonal H3N2 strains whose original form was  responsible for the 1968 influenza pandemic.

We note that FluNet has previously been employed to study the spread of influenza on global or large-scale spatio-temporal patterns in three other studies that we know of.
Finkelman {\it et al.} \cite{Fink+07} studied the pre-2009-pandemic period between January 1997 and July 2006 in 19 temperate countries in both Hemispheres.
They identified large scale co-existence of influenza A and B, interhemispheric synchronized pattern for subtype A H3N2, and latitudinal gradients in the epidemic timing for seasonal influenza A.
A recent study \cite{WPRG12} that focused on the Western Pacific Region between January 2006 and December 2010, found that dominant strains of influenza A were reported earlier in Southern Asia than in other countries.
Thus, status in South Asian countries may provide early warning for other countries.
Bloom-Feshbach {\it et al.} \cite{Bloo+13} examined latitudinal variations in seasonal activity of influenza and respiratory syncytial virus (RSV) and applied a time series model to the seasonal influenza data from 85 countries.
They found evidence of latitudinal gradients in timing, duration, seasonal amplitude and between-year variability of epidemics.
 In terms of the  temporal pattern in a single region,
Dorigatii {\it et al.} \cite{Dori+13} studied the third wave of infection by the H1N1pdm pandemic strain in England in the 2010/11 season. They found that increased transmissibility and loss-of-immunity among the population may be responsible for this unexpected wave.

However, to our knowledge, no study has been conducted focusing on the global pattern of seasonal activities of the H1N1pdm  pandemic virus and interactions among different strains based on the FluNet large-scale dataset from 2010 to 2013. This is of special interest given that the surveillance scale was substantially improved since 2010. Such a study is important to aid the development of strategies for relieving the burden of seasonal influenza. Understanding the spatial pattern may be useful in a global effort to reduce the impact of a deadly influenza pandemic. The activity of H1N1pdm still causes substantial attention in the post-pandemic era and has led to substantial morbidity and mortality in most years since its appearance, including 2013/14.

On the global network of influenza transmission, it is known that China and Southeast Asia lie at the center of the global network and USA acts as the primary hub of temperate transmission \cite{Russ+08,Bedf+10}. The expansion of H1N1pdm during 2009 can be explained with data on human mobility (air travel) and viral evolution\cite{Leme+14}.

\section{Materials and Methods}\label{S:mm}

Weekly time series of lab-confirmed cases (isolates) of influenza were obtained from FluNet
for 138 countries that have non-zero cases between January 2006 and May 2014.  The analysis included six different types of time series: {\bf i)} total specimens processed, {\bf ii)} H1N1pre strains, {\bf iii)} H3N2 strains, {\bf iv)} H1N1pdm strain, {\bf v)} un-subtyped influenza A, and {\bf vi)} influenza B (including two circulation lineages).

The number of un-subtyped influenza A is often substantial and needs to be accounted for. Following \cite{Fink+07}, we proportionally assigned the un-subtyped influenza A to the three subtypes as follows. Let the number of lab-confirmed cases of subtypes H1N1pre, H1N1pdm, H3N2 and un-subtyped A for any country in a particular week be $a_1, a_2, a_3$ and $a_0$, respectively.
Then the new revised number for each  of the three subtypes was taken to be: $a_i'= a_i+a_0a_i/\sum a_i$ for $i=1,2,3$.

The statistical analysis was implemented in the R programming language (http://www.r-project.org/). We generally preferred to focus on regions (macro geographical continental regions and geographical sub-regions) rather than their constituent countries (see Figure~\ref{Fig:region}) because aggregated regional data is less influenced by stochasticity.  We observed that nearly all countries, especially those in the temperate regions, largely followed their regional patterns.
The breakdown of countries of the eight regions used in this study may be found in the supplementary material section~\ref{S:region}.

To help comparisons between hemispheres it was convenient to redefine the initiation and termination dates of calendar years in a manner that makes influenza seasons (usually winter) line up.
We therefore moved forwards the beginning and end dates that define years for Northern Hemisphere (NH) countries to stretch from the 35th week of a calendar year to the 34th week of the following calendar year, roughly overlapping the school calendar year.
For countries in the Southern Hemisphere (SH), the calendar year remains the reference frame. Thus the skip-year (skip-season) of H1N1pdm is 2011/12 in NH and is 2012 in SH, see Fig.~\ref{Fig:skip}.

A skip-year, or simply a {\it skip}, is defined as a season with an uninitiated or minor epidemic.
What constitutes a ``minor" epidemic is difficult to quantify precisely.
For the purposes of this study, we formulated the following practical quantitative comparative definition.  If, after an epidemic year the number of influenza cases drops by more than a factor of ten, we consider this to be a skip year.  We thus use the following  simple measure defined for H1N1pdm:
\begin{equation}\label{E:skip}
\alpha_1 = \log\{(h_{11}+k)/(h_{10}+k)\},
\end{equation}
where $h_{10}, h_{11}$ are the total number of lab-confirmed cases of H1N1pdm in that region during the 2010/11 and 2011/12 seasons, respectively.
The index compares the ratio of the number of cases in 2011/12 season to those in the 2010/11 season in NH (or 2012 to 2011 in SH).
Our criterion for a ``skip" is generally that $\alpha_1 < \log (1/10)$, {\it i.e.}, an order of magnitude difference.
We set $k=50$ in $\alpha_1$ to reduce errors magnified when case numbers are small. The merit of using skip index (which is a ratio of two years) rather than the actual number is clear. In this way, we can remove the differences in the testing effort among countries and we can also remove the effects of different population sizes among countries. We assume that the total numbers of specimens processed had not varied much from year to year, which was true from 2010 to 2012.

Similarly we define skip-indices $\alpha_2,\alpha_3,\alpha_4$, for subtype H3N2, influenza B and total specimens processed, respectively.
We argue that as long as surveillance efforts and testing policies were implemented consistently in each country between 2010 and 2013, then effects due to differences in testing policies are removed by taking the ratio of confirmed cases over consecutive years.
To our knowledge, there were no dramatic changes in surveillance effort in most countries from 2010 to 2013 (as observed from  the total number of specimens processed).

\section{Results}\label{S:results}

\subsection*{Skip-and-resurgence pattern of H1N1pdm}

\begin{figure}[ht!]
\centerline{\includegraphics[width=17cm]{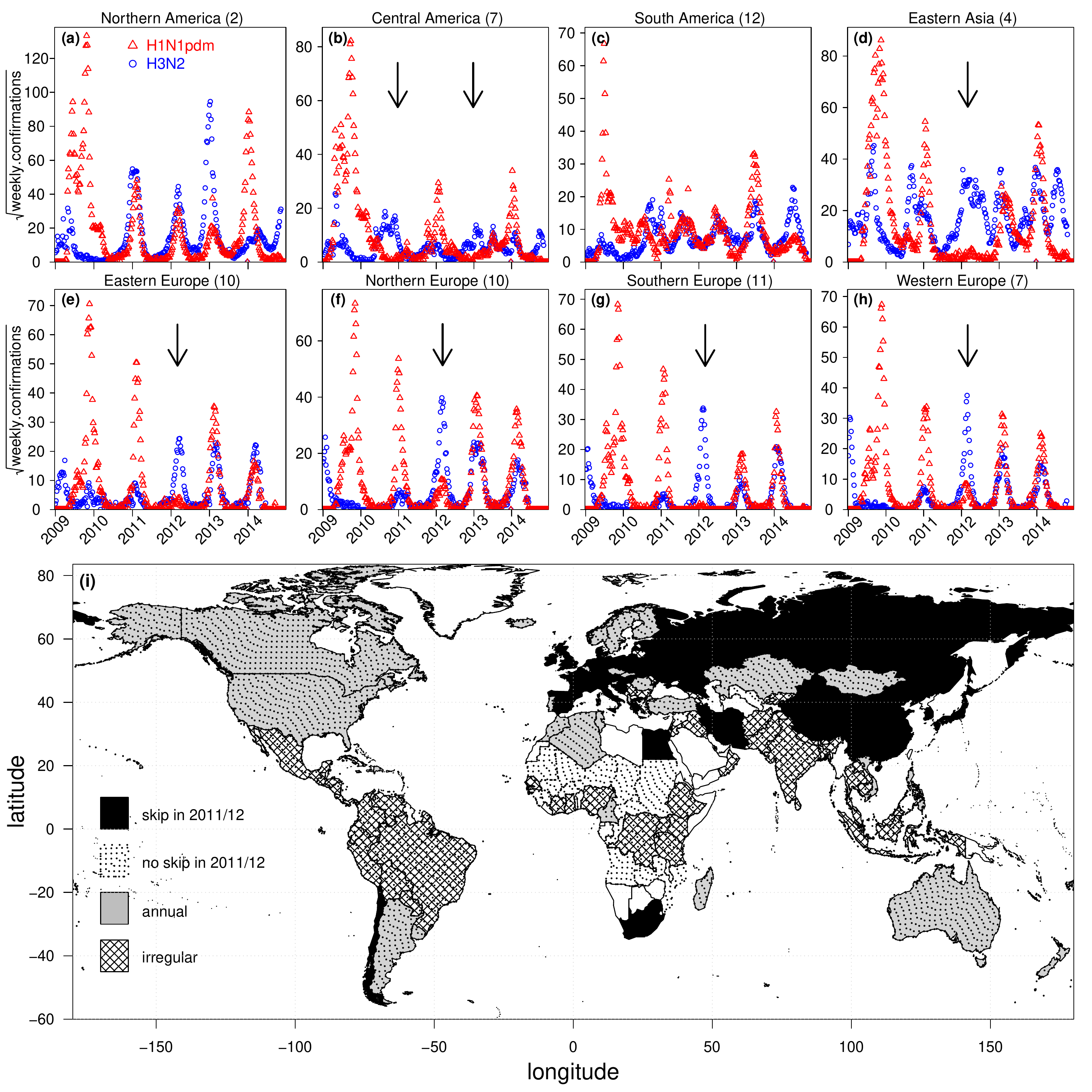}}
\caption{Spatio-temporal patterns of H1N1pdm and H3N2.
(Panels a-h) Square-root of weekly lab-confirmed cases of H1N1pdm (red triangle) and H3N2 (blue circle) in eight geographical regions between January 2009 and May 2014.  Black arrows indicate `skip' seasons for H1N1pdm. Northern America (panel a) exhibits annual epidemics without a skip; Central America (panel b) exhibits seemingly biennial epidemic with a skip in both 2010/11 and 2012/13 seasons; South America (panel c) is irregular in pattern;  Eastern Asia (panel d) and Europe (panel e-h) exhibit annual epidemics with a skip of H1N1pdm and a substantial epidemic of H3N2 during the 2011/12 season. Panel (i) summarises the global pattern during the 2011/12 season.
}
\label{Fig:region}
\end{figure}

Figure \ref{Fig:region} panels (a-h) show weekly aggregated lab-confirmed cases of subtypes H1N1pdm (red triangle) and H3N2 (blue circle) in the eight geographical regions having the largest case numbers in the period January 2009 and May 2014.
A similar set of panels for thirty different countries (having the largest number of confirmed cases) may be found in Fig.~\ref{Fig:individual} in the supplementary material. It should be emphasised that the graphs are scaled to highlight the trends (and also accommodate the extremely high peak in 2009) by plotting the square-root of the weekly lab-confirmed cases. The figures immediately identify a number of clear features. With regard to H1N1pdm, we summarise here:

\begin{itemize}
\item [(i) ] All regions in Europe and Eastern Asia have identical  trends and experienced skip-years in 2011/12. In more detail these regions experienced two initial waves of H1N1pdm in 2009/10, followed by a single wave in 2010/11, a skip-year  in the 2011/12 season and then a reemergence of H1N1pdm in the following 2012/13 season.  The skip was more evident in Eastern/Southern Europe than Western/Northern Europe. Although the latter experienced a  minor  outbreak in 2011/12, its peak was an order of magnitude lower than the previous season, and thus by our criteria could be classified as a skip. The size of the mini-outbreak is misrepresented and appears exaggerated due to the square-root scaling.

\item [(ii)] In stark contrast, H1N1pdm failed to show any skip in Northern America.  In fact H1N1pdm exhibited annual oscillations in Northern America, with an early and large wave appearing in the 2013/14 flu season.

\item [(iii)] Central America, where H1N1pdm originated, shows a different pattern to that of Europe and Eastern Asia. Instead skips occurred both in 2010/11 and 2012/13  but not in 2011/12  (Fig.1b).
The dynamics over these years were essentially biennial.

\item [(iv)] South America shows an irregular pattern (Fig.1c).

\end{itemize}

With regard to H3N2 dynamics, we observe:

\begin{itemize}
\item [(i)] All regions in Europe and Eastern Asia experienced significant H3N2 epidemics  in 2011/12,  which was a skip-year for H1N1pdm.  Moreover, apart from 2012-2014, the H3N2 dynamics were essentially negatively correlated with H1N1pdm.
\item [(ii)] In Northern America H3N2 tended to oscillate synchronously in-phase with H1N1pdm in 2010-2012.
\end{itemize}
A spatial summary of the dynamics of each geographic region has been superimposed on the world map of Fig.~\ref{Fig:region} panel (i).
The regions colour coded in black experienced a skip year in 2011/12 and constitute a considerable proportion of the global map.

%
%

\begin{figure}[ht!]
\centerline{\includegraphics[width=16cm]{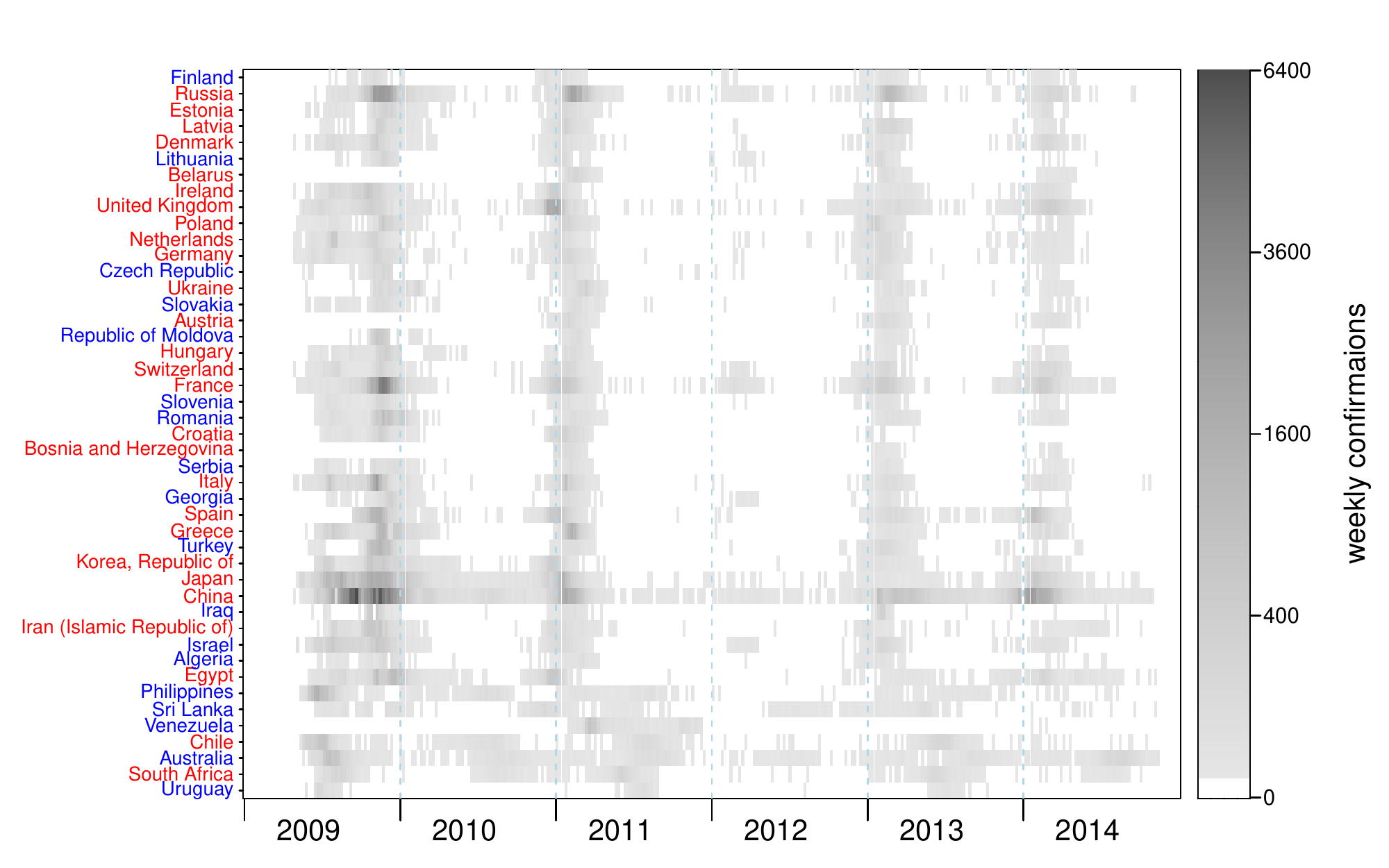}}
\caption{Countries where H1N1pdm skipped the 2011/12 season (ordered in latitude order from North to South). Countries with $\alpha_1 < \log 1/10$ are coded in red and countries with  $\log 1/10< \alpha_1<\log 1/5$ are in blue.}
\label{Fig:skip}
\end{figure}

We identified 27 countries with $\alpha_1 < \log(1/10)$ and thus skip years. Total of confirmations of the skip year was one order of magnitude lower than that of the previous year. Using a higher threshold $\alpha_1 < \log(1/5)$, the number increases to 45 countries. Namely 18 countries have $\alpha_1$ between $\log(1/10)$ and $\log(1/5)$. The weekly confirmations of these latter countries are displayed in Fig.~\ref{Fig:skip} (in latitude order) which gives a remarkable demonstration of the broad geographic synchrony of the epidemic skip over the globe. We note that most of the 45 countries experienced a resurgence of H1N1pdm in 2012/13.

\subsection{The 2011/12 skip year and strain dynamics}

Fig.~\ref{Fig:skip} makes clear how the influenza dynamics of many countries are strongly correlated in time and skip synchronously in the 2011/12 period.
The time series in Figs.~\ref{Fig:region} and \ref{Fig:individual} suggest that for nearly all countries the H1N1pdm and H3N2 cases are negatively correlated over the full period Jan 2006 to May 2014.
This is exemplified in the 2011/12 season where H1N1pdm skipped in most countries while H3N2 outbreaks occurred in its place. To study this relation in more detail, we tested whether these two strains are correlated across all countries  in the 2011/12 season alone. That is, we asked whether countries with smaller H1N1pdm outbreaks tend to have larger H3N2 outbreaks, in 2011/12 in NH (or 2012 in SH).
Using the skip-index, eqn.~\ref{E:skip}, our analysis showed that of the 108 countries which reported more than 500 cases of all strains, the correlation coefficient across all countries is $r=-0.61$. (Without the threshold of 500, the correlation is $r= -0.63$.)

For a more detailed analysis of the 2010-2012 years, we  considered a linear model with $\alpha_1$ as the response, and having seven different predictors: $\alpha_{2}$ (H3N2), $\alpha_{3}$ (flu B), rank of population size in the year 2005, rank of area, rank of absolute latitude, rank of distance from Mexico and  geographical region code. Regarding the distance from Mexico, we considerd both Euclidean distance (defined as $\sqrt{x^2+y^2}$ where $x,y$ are in terms of longitude and latitude, respectively) and effective distance \cite{Broc+13} and found no significant difference.

\begin{equation}
\alpha_1 =  c_1 + c_2 \alpha_2 + c_3 \alpha_3 +c_4 \text{popn.rank} + c_5 \text{area.rank} + c_6 \text{dist.rank} ...
\end{equation}
where the $c_i$ are constants to be fitted.

The H3N2 skip index ($\alpha_{2}$) was found to be a significant predictor ($p$-value $<0.001$) of $\alpha_1$ though negatively correlated,
while $\alpha_3$ (influenza B) was not a significant predictor ($p$-value $\approx 0.452$). Region code and area rank were also found to be significant predictors ($p$-value $<0.001$ and $\approx 0.01$ respectively), while all other predictors were not significant. These results parallel our observations that in broad terms, countries in the same region share a common pattern.

%
%

In this study, we have focused largely on the 2011/12 skip. However, it is evident from Figure 1 of \cite{Fink+07} and Fig.\ref{Fig:spatialpre} in the supplementary material that H3N2 exhibited a similar skip in 2000/01. After obtaining FluNet data for the period between 1995 and 2005 (see supplementary material section~\ref{SI9505}) we repeated the above analysis. The H3N2 skip-index for 2000/01 season was found to be negatively correlated with both H1N1pre $(r=-0.407)$ and influenza B $(r=-0.573)$ across 72 countries.  With a generalized linear model (the skip-index of H3N2 as the response and these of H1N1 and B and countries absolute latitude as factors) both H1N1 and B were significant ($p-$value $<0.001$). New variants of both H1N1pre and influenza B emerged in 1999 (A/New Caledonia/20/99 and B/Sichuan/379/99, respectively), which possibly played a role in the skip of 2000/01  for H3N2 uniformly across all countries.  It is worth noting that A/New Caledonia/20/99 had been in the vaccine components for seven seasons. The new variant of H3N2, which may have enhanced the 2011/12 skip of H1N1pdm in Europe and Eastern Asia, was most likely A/Perth/16/2009 (\url{http://www.hpa.org.uk/webc/HPAwebFile/HPAweb_C/1226908885446}).

\subsection{Mathematical Model}

 We made use of modern mathematical modelling techniques \cite{He+10} to fit a stochastic single-strain Susceptible-Exposed-Infectious-Recovered model ($SEIR$) to the FluNet influenza data from 2009 until the end of 2013. Details of the model are given in the supplement
 section \ref{s_model}. The original goal was to understand better those factors that caused the 2011/12 skip.
The model fits were made  for the ten countries having the largest total confirmations since the invasion of the strain in 2009.

 The following assumptions were made when fitting the model:
 \begin{itemize}
 \item The initial susceptible proportion of the population was taken to lie between 40\% and 70\% for all countries rather than 100\%. This takes into account that many of the elderly population had preexisting cross-reactive antibodies\cite{Hanc+09}. In addition it was found that the cross-protection provided by the pre-pandemic vaccine was as high as 19\% \cite{Yin+12}.

\item The transmission rate $\beta(t)$ was taken to be seasonal and  modelled by a periodic function of time, with a period of one year. Weather variations and school terms are understood to be responsible for the seasonal variability \cite{Yaar+13,Sham+10}. We adopted a seven-node cubic spline function, and fixed the parameter of node seven to be equal to node one. The function is second-order differentiable except for the seventh node. Thus there were six free parameters in transmission rate $\beta(t)$.

\item The reporting rate $\rho(t)$ of each country was modelled by a three-piece step function of the following form:
    \begin{equation}
    \rho(t)=\begin{cases}   \rho_1,& t \in [\mbox{2009-1-1, 2009-6-11}]\\
    \rho_2,& t \in (\mbox{2009-6-11, 2009-8-31}]\\
    \rho_3,& t \in (\mbox{2009-8-31, 2014-2-28}]\\
    \end{cases}
\end{equation}
Here, 2009-6-11  is the date WHO announced the initiation of the 2009 pandemic and 2009-8-31
is the end of the 2008/09 flu season and the start of 2009/10 flu season.
This allows for the sudden increase of the reporting rate during the  2009 pandemic.
 For example,  the reporting rate changed dramatically during 2009 in the UK \cite{Dori+13} and in Canadian provinces\cite{Earn+12,He+13b}.

\item If the infection dies out in a country after the invasion in the simulation, we introduced a single infected individual. This mimics the transmission of influenza between countries, so that no country is completely isolated.

\item The latent period ($\sigma^{-1}$) of the disease was assumed to to be one day and the infectious period ($\gamma^{-1}$) two days.

\item We also fit the duration of the immunity ($\lambda^{-1}$) by calculating the maximum log likelihood profile for it. Namely we fixed the duration to eight discrete values spanning from 1.5 to 7 years, and maximized the performance of the model while fitting other parameters. Then from this profile we estimated $\lambda^{-1}$, and its 95\% confidence interval \cite{He+10}.

\end{itemize}

\begin{figure}[h!]
\includegraphics[width=17cm]{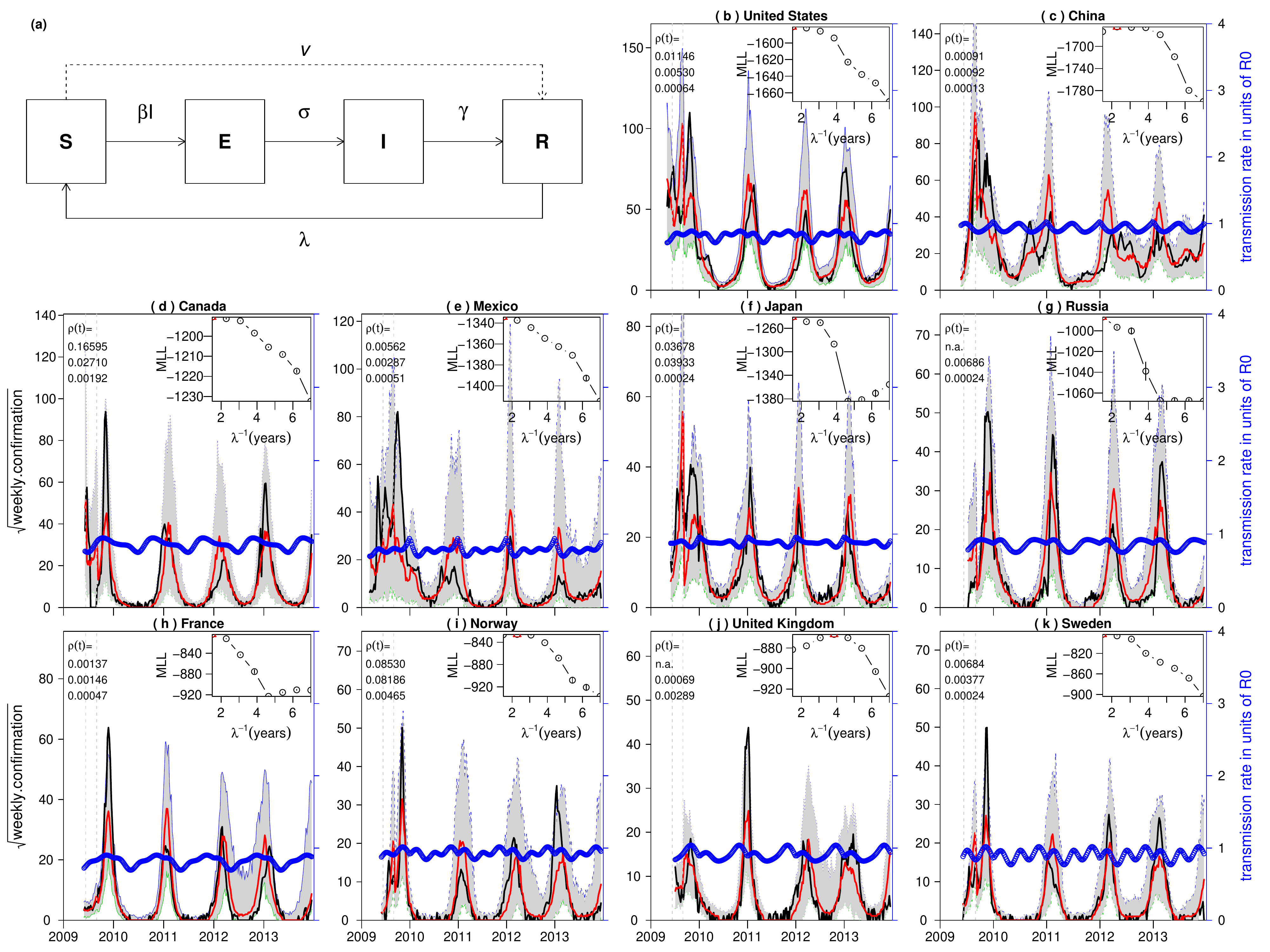}
\caption{Fitting an SEIR model to influenza A confirmations in 10 countries. Panel {\bf a}, a flowchart of the model. Panel ({\bf b-k}), the results in 10 countries. The inset panel shows the profile log likelihood for the duration of immunity.
Each panel shows the simulation (red) versus the observed (black), with the best fitting parameters.
The blue curve shows the estimated transmission rate in the unit of basic reproductive number. The dotted vertical lines indicate the two timings for the reporting ratio changes. The simulations are median values for each week of 1000 simulations and shaded region show the 95\% range.
}\label{Fig:fitting}
\end{figure}

The model essentially finds the best fitting estimates of the transmission rate and reporting ratio (as defined  above) to the influenza A time series by  maximizing the relevant log likelihood.
The output of the model is a plot of the profile log likelihood as a function of the duration of immunity.  For example, Fig.\ref{Fig:fitting} shows the best fitting model to the FluNet time series data (plotted in black) from ten countries when aggregating influenza  A ({\it i.e.}, by combining H1N1/99 data with H3N2.
The inset figure plots the likelihood profile and shows that the maximum occurs when the immunity duration ($\lambda^{-1}$) is approximately two years in most countries, which is biologically reasonable \cite{Cowl+14}.

We plot time series of the median value of reported cases for 1000 model simulations.  The median values are plotted in red, while the grey shaded region indicates the 95\% confidence interval.
The median values sit close to the observed values (black lines) for all ten countries.

The fitted reporting ratio is shown in the top-left corner, and the estimated transmission rate is shown as a solid blue curve. The simulations match most of the observed waves for all ten countries. The estimated seasonal amplitudes in the transmission rate are small and largely consistent across countries, which suggests that the post-pandemic waves are largely due to  loss-of-immunity and its associated replenishment of the susceptible pool (possibly impacted by dynamical resonance \cite{Dush+04}). The estimated reporting ratio is very small and varied considerably across countries. Recall that in \cite{Dori+13}, it was found that only 0.7\% infected individuals actually visit a General Practitioner.
If we accept Dorigatti et al's  \cite{Dori+13} estimation of GP visits (0.7\%) and examine all ten countries, we find that between 1/50 to
2/3  (median 1/14) of GP visits of infected individuals led to positive lab-confirmations, depending on the country in the post-pandemic period.

We also attempted to use the same model for fitting  single strain data including H1N1pdm alone and H3N2 alone.  However, when modelling a single strain (either H1 or H3), fits were generally poor.
In particular, the model was unable to predict the skip in 2011/12 for any country.
Since the main shortcoming of the model used is that it is only single strain, we conclude that a multi-strain model that includes the interaction between  the H1N1pdm and H3N2 strains is needed to capture the 2011/12 skip.

\subsection*{Synchrony Patterns}

We examined the synchrony pattern across Northern Hemisphere (NH) countries (latitude $> 29^\circ$) for the three strains, H1N1 (combined H1N1pdm and H1N1pre), H3N2,  and influenza B. We focused on the period from Jan 2006 to May 2014.

In order to quantify synchrony, following \cite{Fink+07} we made use of the Mean Confirmation Time index, or MCT.
The MCT for country-$j$, is the mean time of infection of all infected persons over the course of the epidemic under examination. Thus, if in country-$j$ there are  $n_i$ confirmations in the $w_i$\,th week of a season for a particular strain, then

\begin{equation}
\text{MCT}_j=\sum\limits_i n_iw_i/\sum\limits_i n_i.
\end{equation}

\begin{figure}[h!]
\centerline{\includegraphics[width=16cm]{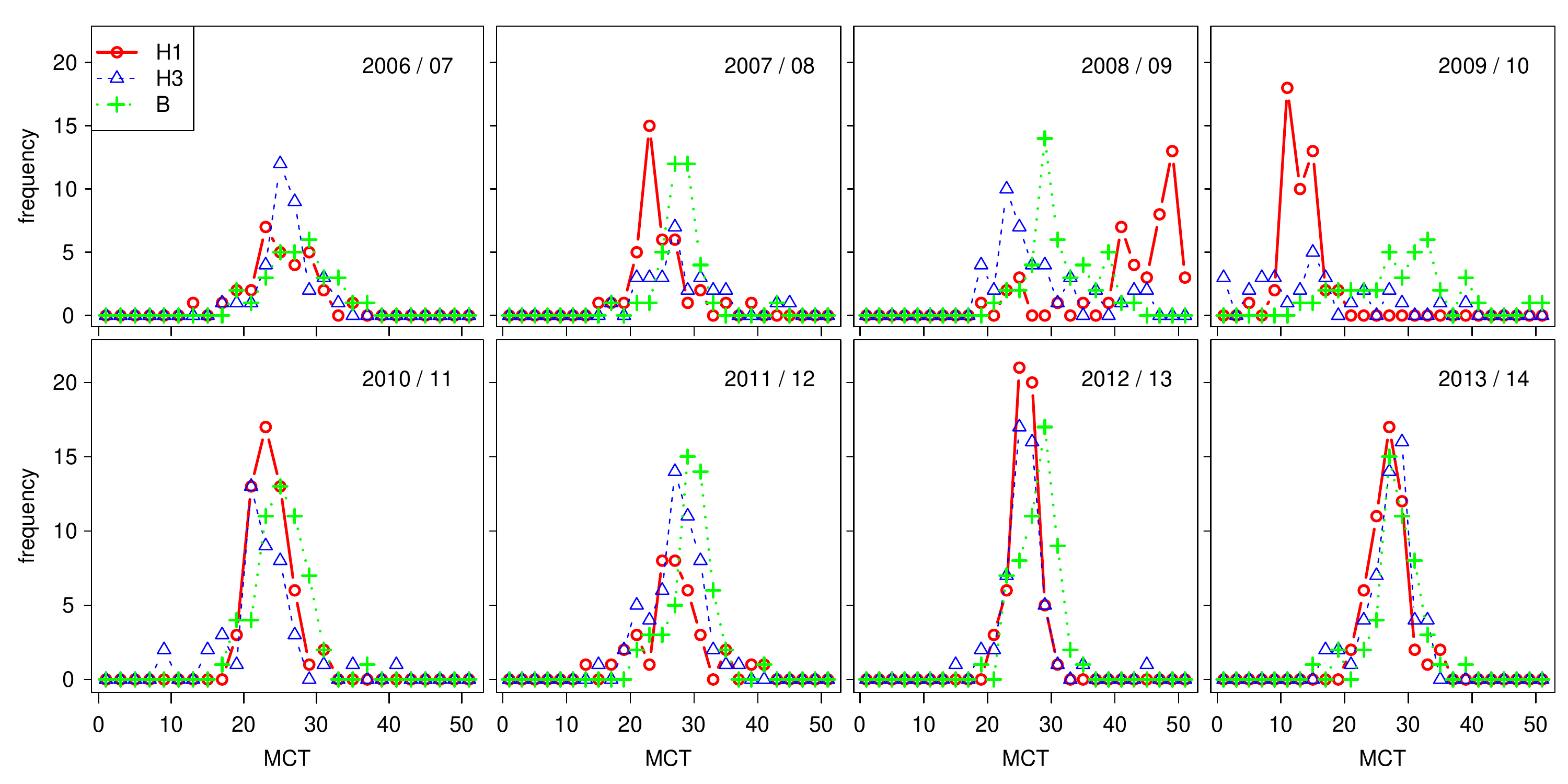}}
\caption{Distribution of Mean Confirmation Time for three strains, H1N1 (H1N1pre and H1N1pdm), H3N2 and Influenza B, in North Hemisphere in eight flu seasons.}
\label{Fig:synchrony2}
\end{figure}

If all countries have the same MCT, so that $\text{MCT}_j=c$ is the same constant for each country-$j$, then the distribution of the $\text{MCT}_j$ is just a single spike indicating that the countries are highly synchronized.
 Obviously, the smaller the standard deviation amongst the different $\text{MCT}_j$ the more synchronized are the different countries.
The synchrony analysis examines the distribution of $\text{MCT}_j$ over all countries $j=1,2,..., N$ (median and standard deviation) for different strains and different seasons. Countries with no cases were removed from the analysis. The results are shown in Fig. \ref{Fig:synchrony2} and listed in Table \ref{synchrony}.
We examined the NH countries and for this purpose grouped H1N1pre and H1N1pdm together as H1N1.

Fig.~\ref{Fig:synchrony2} show that countries were  more synchronized by H1N1pdm than by H3N2 or influenza B in 2010/11 and 2012/13. Indeed, from Table \ref{synchrony} in the supplementary material, the standard deviation of MCT for the H1N1pdm strain in the 2010/11 and 2012/13  seasons were significantly smaller than the other seasons and any of the other strains.
This may reflect a more efficient transmissibility of the H1N1pdm virus which allows it to spread more rapidly between countries. These findings corroborate what is observed by eye in Fig.~\ref{Fig:skip} and Fig.~\ref{Fig:spatial}, and that the MCT of H1N1 only varies by some 5 weeks across all countries.
 Note that the median of MCT of influenza B seems larger than the two influenza A strains (by two weeks), suggesting that the flu B epidemic lagged behind the other two flu A strains \cite{Fink+07}.

\subsection*{Patterns of Strain Replacement}
\begin{table}[h!]
\begin{center}
\caption{Annual Total Confirmations of H1N1pre and Total Specimens Processed}\label{H1N1pre}
\begin{tabular}{cccccccccc}
\hline
year & 2006 & 2007&  2008&  2009&  2010&  2011&  2012&  2013&  2014 \\
H1N1/07 & 13268 &18983& 29807& 37879&   709&    41&     3&     5&    17 \\
Specimens& 355834 &  513430 &  671232 & 2290733 & 1186197 & 1270287 & 1350542 & 1672204 &  810570\\
\hline
\end{tabular}
\end{center}
\end{table}
Table \ref{H1N1pre} lists the annual total confirmations of H1N1pre.  It is interesting to note that the  2008 total (pre-pandemic)
was double that in 2006, which was due to an increase in testing effort (total specimens processed).
The high number in 2009 was most likely due to the extensive testing during the pandemic.
The numbers decreased quickly after 2009 suggested that H1N1pre was replaced by H1N1pdm \cite{Pica+12}
The low numbers in 2012-2014 are likely to be errors, either misclassification or mis-input. For example, six cases of H1N1pre were
reported in Poland in the 7th week of 2014. However, close observation revealed that there were minor epidemics of both H1N1pdm
and H3N2 in that period, yet data was unexpectedly  missing in these categories.  But despite its presence anywhere else, six cases of H1N1pre were
recorded suggesting possible misdiagnosis. No evidence for an epidemic of H1N1pre after 2011 has been found so far.
It is interesting to note that the original form of the H1N1pre subtype had an unusual  re-emergence in 1977 some 20 years after its  disappearance\cite{Wert10}.

\section{Discussion}

Fundamental epidemiological principles are able to explain the skip dynamics seen in Figs.~\ref{Fig:region} and \ref{Fig:skip} in relatively simple terms.  When the new strain H1N1pdm first appeared in March 2009, the population at large had no previous exposure to the strain.  This allowed the pandemic to develop into a global-scale epidemic even though outside the normal influenza season in many countries.  With the passage of time, each successive epidemic outbreak exposed the population at large further to the new H1N1pdm strain, thereby building up population immunity and reducing  the number of susceptible individuals
\cite{Dori+13}.  Thus by the end of 2011/12,   the susceptible pool of individuals available for infection had reduced below a critical threshold level, so that the epidemic failed to trigger over the 2011 winter season.  In epidemiological parlance, by ``burning out" the available susceptible pool, the virus effectively reduced the reproductive number \Ro below unity making it impossible for the epidemic to initiate in the 2011/12 season. This set the stage for the opportunist H3N2 virus to outcompete and replace H1N1pdm, thus accounting for the H3N2 outbreak at the end of  2011.  Fig.\ref{Fig:region} makes clear the complex interaction between the  H1N1pdm and H3N2 strains as they compete for the available pool of susceptible individuals as well as offer cross-protection.

 It is interesting that in Central America where H1N1pdm first appeared, the outbreak progression was different to the above pattern. Two skip-seasons were observed in 2010/11 and 2012/13  (see Mexico in Fig.~\ref{Fig:individual}). These skips were in all likelihood an outcome of the same underlying process, namely a burn-out of susceptibles from the previous waves.

Other mechanisms such as climatic variation, poor surveillance and results of new births unlikely played a key role here. There were no previous studies showing that these factors favor H3N2 rather than H1N1pdm. These factors are largely the same between Europe and Northern America.

%
%

The occurrence of skips gives information regarding the loss-of-immunity (strain specific) in the population, particularly if there might be only a single viral strain, or if the viral strain is stable and evolves only at a relatively slow rate.  The latter is the case for
the H1N1pdm strain which is believed to have been  antigenically stable since its emergence in 2009 \cite{Guar+13}.
As an indication of its stability, the vaccine component against H1N1pdm recommended by the WHO and the United States Food and Drug Administration (FDA) was not updated since fall of 2009, while those vaccine components against H3N2 and influenza B have been updated more than twice already (see \ref{s_component}).  If the H1N1pdm strain was indeed stable over these last years,  and the  virus evolved relatively slowly, then the main source of new susceptibles in the population was largely derived  through natural loss of immunity. In these circumstances,
the resurgence of H1N1pdm in 2012/13 after the skip in 2011/12 should be viewed as a consequence of the natural loss-of-immunity in the population \cite{Cowl+14}.

The differences in influenza dynamics between Northern America (no skip) and Europe (skip) given that they share many common factors  with regard to economics, culture, climate and latitude are in some respects surprising.  We speculate  the different dynamics may be connected with the influenza vaccination coverage  which was consistently higher in Northern America than in Europe (and the rest of the world). High coverage of vaccination (against H3N2) among general population could have slowed down the transmission of H3N2, thus saved H1N1pdm from skip a year in Northern America. Vaccination coverage has been consistently close to 40\% in the United States and 30\% in Canada, but less than 30\% in Europe (see supplementary material section \ref{s_vaccine}) and other parts of the world, for example 14\% in Hong Kong \url{(http://www.chp.gov.hk/}) (see the skip in Hong Kong in \ref{Fig:individual}).  Also intriguing is that many parts of Northern America and Central America  had much higher attack rates of H1N1 in 2009 and influenza-associated mortality in 2009 was almost 20-fold higher in some countries in America than in Europe (see \cite{Simo+13}). Additional work is still needed to understand which factors are responsible for the different spatio-temporal patterns of influenza seen in Europe and America.

Our attempts to fit the time series of aggregated influenza A confirmed cases using the same simple SEIR model are  shown in Fig.~\ref{Fig:fitting} and are reasonably well given that they capture the different trends observed in ten different countries.
That the same SEIR model reproduced the different trends suggests that the dynamics of influenza epidemics have a large degree of determinism and the model is considerably robust.
Moreover this indicates that the
the key assumptions behind the SEIR model are largely being met.  Namely, the classical mathematical concept of infection being spread by a  randomly mixing population and mean field dynamics appear to apply when modelling large real human populations.  The different features of each country's influenza A time series can be explained through a change in the SEIR  model's  parameters.
It is also interesting that the model fits (likelihood profiles) indicate a reasonably fast loss of immunity in the vicinity of 2-4 years. This could explain the fast susceptible buildup required during skip years, which would be necessary to generate the resurgent epidemics observed in the following years.

Our analysis has given interesting insights into the global patterns of the invasion of H1N1pdm, which first appeared as a pandemic and then, within a few years,  apparently outcompeted and completely replaced the H1N1pre seasonal flu strain.  The synchrony between countries of the H1N1pdm outbreaks is striking particularly as witnessed in the highly visible skip (Fig.~\ref{Fig:skip}), where for a large number  of countries, H1N1pdm failed to outbreak in the 2011 flu season.  Moreover, the synchrony between countries in H1N1pdm outbreak years was also considerably strong (see Fig.3).
The FluNet data gave a comprehensive picture of these phenomenon as they evolved in time over several years, and also in space over 138 countries.

\section{Acknowledgments}
We thank David Earn, Jonathan Dushoff, Ben Bolker and Raluca Eftimie for helpful discussions. DH was supported by a start-up grant from the Department of Applied Mathematics at Hong Kong Polytechnic University, a``Central Bidding'' grant from Hong Kong Polytechnic University, a RGC/ECS grant from Hong Kong Research Grant Council, and a Health and Medical Research Grant from Hong Kong Food and Health Bureau Research Council.

\setcounter{figure}{0}
\setcounter{table}{0}
\setcounter{section}{0}

\makeatletter
\renewcommand{\thetable}{S\@arabic\c@table}
\renewcommand{\thefigure}{S\@arabic\c@figure}
\renewcommand{\thesection}{S\@arabic\c@section}

\clearpage
\centerline{\huge \bf Supplementary Materials}

\section{Synchrony Pattern}

\begin{figure}[h!]
\centerline{\includegraphics[width=12cm]{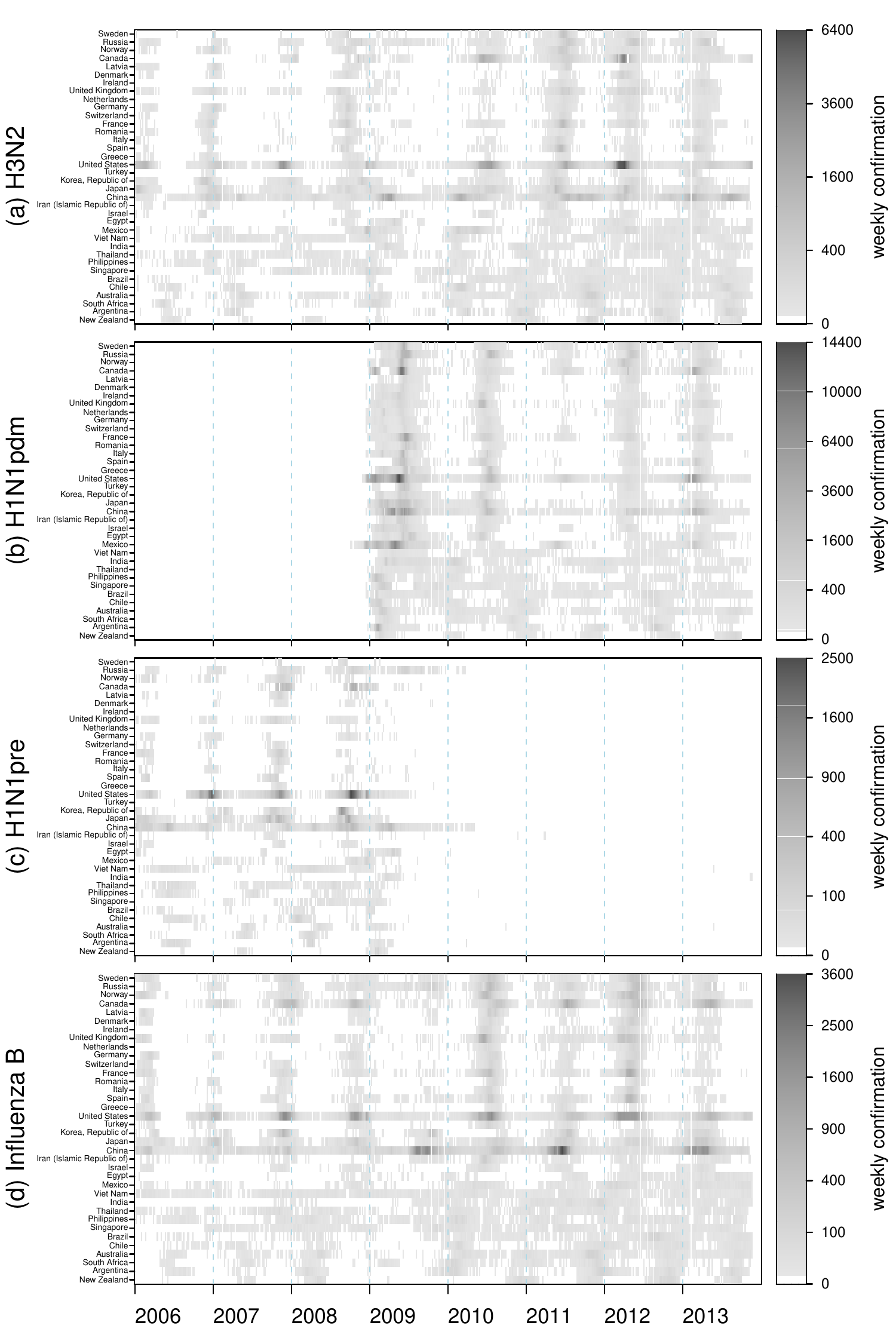}}
\caption{Spatio-temporal patterns of lab-confirmed cases of four types of influenza strains from 35 countries  between January, 2006 to May, 2014. Countries are listed from north to south. Grey scale shows the weekly lab-confirmed cases. The horizontal axis is time in weeks. Evident synchrony patterns for H1N1pdm, H3N2 and influenza B can be seen in the temperate countries. H1N1pdm showed skip in most countries in Eastern Asia and Europe during the 2011/12 season. H1N1pre was replaced by H1N1pdm after the 2009 pandemic. Isolated cases of H1N1pre after 2011 are likely due to errors.}
\label{Fig:spatial}
\end{figure}

Figure \ref{Fig:spatial} shows the spatio-temporal patterns of weekly lab-confirmed cases of four types of influenza strains from 35 countries chosen for having the largest total number of cases between January, 2006 and May, 2014. Panels (a) to (d) from top to bottom show the patterns of H3N2, H1N1pdm, H1N1pre, and Flu B, respectively. Countries were listed according to their latitudes from north to south (data from \url{http://thematicmapping.org/downloads/world_borders.php}).

\begin{table}[h!]
\begin{center}
\caption{Synchrony Indicators}\label{synchrony}
\begin{tabular}{cc|ccc}
\hline
 Indicator & Season  & H1N1 & H3N2 & Flu B \\
\hline
\multirow{6}{*}{ Median of MCT }  & 2006 - 07 & 25.2  & 25.9  & 27.9  \\
& 2007 - 08 & 23.7  & 27.0  & 27.8  \\
& 2008 - 09 & 47.1  & 25.5  & 29.7  \\
& 2009 - 10 & 13.0  & 14.8  & 29.2  \\
& 2010 - 11 & 23.4  & 22.4  & 25.0  \\
& 2011 - 12 & 27.4  & 27.9  & 29.6  \\
& 2012 - 13 & 25.8  & 25.8  & 28.1  \\
& 2013 - 14 & 27.2  & 27.7  & 28.2  \\
\hline
\multirow{6}{*}{ Standard Deviation of MCT }  & 2006 - 07 & 4.39  & 3.32  & 4.48  \\
& 2007 - 08 & 4.21  & 6.23  & 3.73  \\
& 2008 - 09 & 8.51  & 7.51  & 5.05  \\
& 2009 - 10 & 2.47  & 9.80  & 8.41  \\
& 2010 - 11 & 2.54  & 5.53  & 3.51  \\
& 2011 - 12 & 5.59  & 4.04  & 3.67  \\
& 2012 - 13 & 2.14  & 4.02  & 3.09  \\
& 2013 - 14 & 2.97  & 3.73  & 4.01  \\
\hline
\multirow{6}{*}{ Total Confirmations }  & 2006 - 07 &  16011  &  25747  &  11250  \\
& 2007 - 08 &  24325  &  22431  &  28564  \\
& 2008 - 09 & 136679  &  35955  &  23299  \\
& 2009 - 10 & 387667  &  14778  &  28678  \\
& 2010 - 11 & 103534  &  59937  &  56730  \\
& 2011 - 12 &  10014  &  84767  &  45449  \\
& 2012 - 13 &  59003  & 108912  &  69134  \\
& 2013 - 14 & 122365  &  40099  &  42082  \\
\hline
\multirow{6}{*}{ Number of Countries }  & 2006 - 07 &  30  &  34  &  30  \\
& 2007 - 08 &  40  &  28  &  38  \\
& 2008 - 09 &  47  &  42  &  45  \\
& 2009 - 10 &  48  &  30  &  39  \\
& 2010 - 11 &  55  &  44  &  54  \\
& 2011 - 12 &  37  &  55  &  51  \\
& 2012 - 13 &  56  &  53  &  56  \\
& 2013 - 14 &  53  &  54  &  48  \\
\hline
\end{tabular}

\end{center}
\end{table}

From Table \ref{synchrony}, the standard deviation (sd) of MCT for the H1N1pdm strain in the 2010/11 and 2012/13 seasons are significantly smaller than the other seasons and any of the other strains.  In particular, not only the sd of MCT for H1N1pdm in the 2010/11 season is small, but also the total confirmation is high and the number of countries involved is large. Thus globally the 2010/11 wave of H1N1pdm was very substantial and well-synchronized.
Thus the countries are more synchronized by H1N1 than by H3N2 or influenza B. This may reflect a more efficient transmissibility of the H1N1pdm virus which allows it to spread more rapidly between countries. These findings corroborate what is observed by eye in Figure 2. Note that the median of MCT of flu B seems larger than the two flu A strains (by two weeks), suggesting that the flu B epidemic lagged behind the other two flu A strains \cite{Fink+07}.

\section{Spatio-temporal Plots of Individual Countries}\label{s_countries}

\begin{figure}[ht!]
\centerline{\includegraphics[width=16cm]{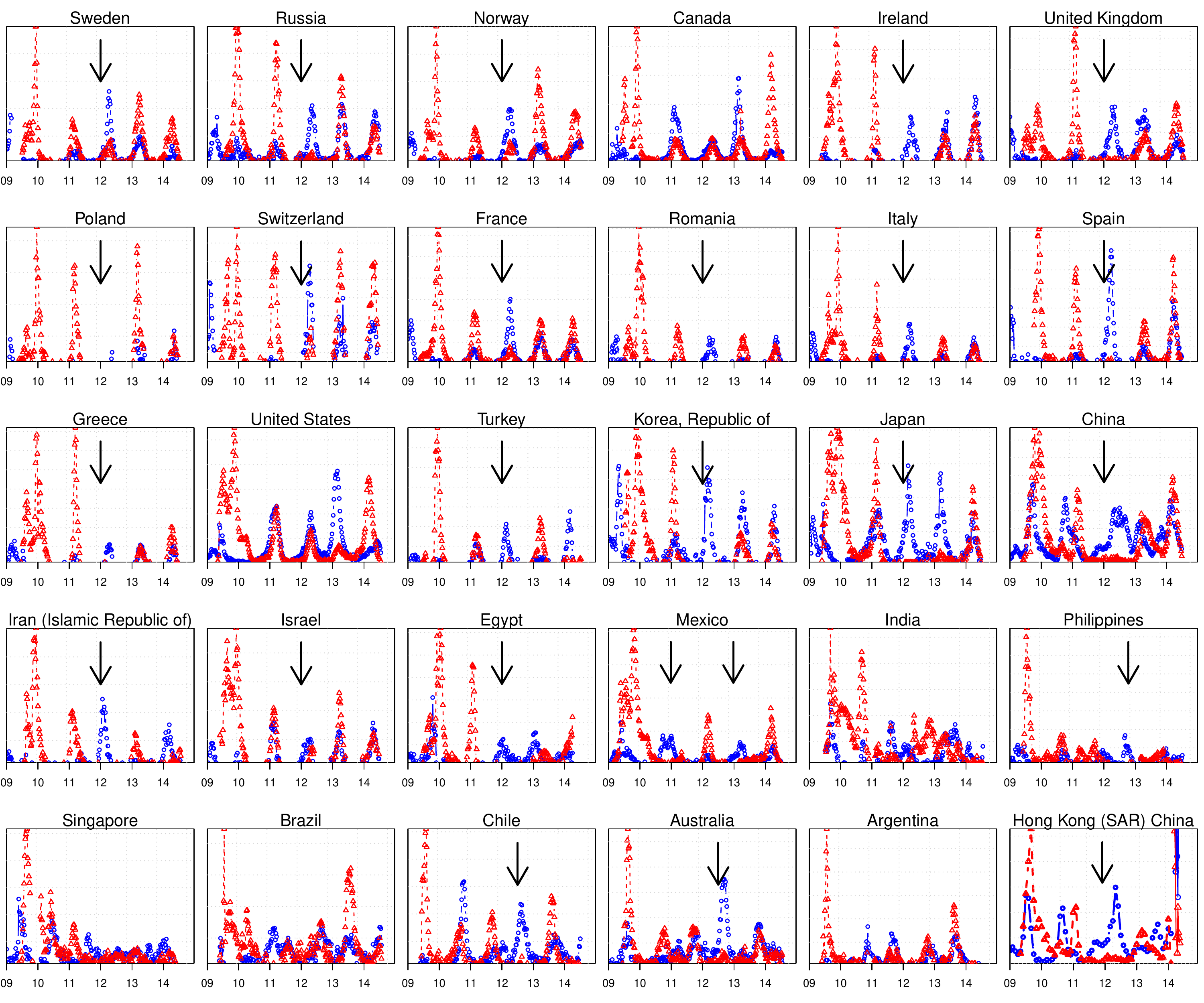}}
\caption{Weekly lab-confirmations of H1N1pdm and H3N2 in 30 populations.
Black arrows indicate `skip' seasons for the  H1N1pdm, when the weekly confirmations of H1N1pdm were evidently low or absent.}
\label{Fig:individual}
\end{figure}

Figure~\ref{Fig:individual} shows individual plots of 30 populations with the largest number of lab-confirmed cases of H1N1pdm and H3N2 between January, 2009 and May, 2014.

There are several countries, such as Japan, which skipped both 2011/12 and 2012/13 season. Australia did show sign of skip in the 2012 season. Its skip index is between 1/10 and 1/5. The vaccination coverages in the general population in Australia and Japan were low.

\section{To Find Countries with Regular Annual Patterns}\label{s_fft}
We consider the weekly influenza A lab-confirmations for each country from 1997 to 2013. If there were zeros in the first few years of the time series, these zero time points will be removed. All NA's in the middle are treated as zeros. Then we apply {\it spectrum} function in R (version 2.15.2) to the time series. We set {\it span}=(3,5), which are the widths of the modified Daniell smoother. We set other parameters at default values (such as {\it taper}=0.1). We compare the heights of the spectrum at frequency $f=1$ and $f<1$, if the height at $f=1$ is at 1.75 times of the mean level of $f<1$, we conclude the time series show regular annual patterns.

\section{Mathematical Model and Likelihood Functions}\label{s_model}

We consider a compartmental (Susceptible-Exposed-Infectious-Recovered) model in the main text,

\begin{subequations}
\begin{eqnarray}
\dot{S} &=& \lambda R - \beta(t)SI - v(t) S\\
\dot{E} &=& \beta(t)SI - \sigma E\\
\dot{I} &=& \sigma E - \gamma I \\
\dot{R} &=& \gamma I - \lambda R + v(t) S
\end{eqnarray}
\end{subequations}
where $S$, $E$, $I$, and $R$ denote the proportions of susceptible, exposed, infectious and recovered proportions in a population, and $S+E+I+R=1$; In particular,
Exposed refers to those infected but not yet infectious;
$\lambda$, $\beta(t)$ and $v(t)$ denote rates of individuals moving from Recovered to Susceptible (due to loss-of-immunity), the transmission rates
of the virus, and the vaccination rate (individuals moving from Susceptible to Recovered, after taking account into the effectiveness of the vaccine and the delay between the vaccination and being effective of the vaccine). $\sigma$ (and $\gamma$) denotes the rates of individuals
moving from Exposed (and Infectious) to Infectious (and Recovered).

It would be very challenging to disentangle $\beta(t)$, $\lambda$ and $v(t)$, given that we assume $\beta(t)$ is time-varying. Mathematically, $v(t)$ can be transferred into as a variation in the $\beta(t)$ \cite{Earn+00a}. But we can roughly estimate the $v(t)$ from other sources (e.g. number of vaccines delivered and the time of the vaccination etc). For simplicity in this manuscript, we set $v(t)=0$, and leave this challenging task to future works.

The weekly report cases is a weekly integral in the form
\begin{equation}
Z_t=\int\limits_\text{a\,week}\rho(t)\gamma Idt
\end{equation}
We assume that the observed weekly lab-confirmations $C_t$ is a random sample from a Negative-binomial (NB) distribution
\begin{equation}
C_t \sim \text{NB}\left(n=\frac{1}{\tau}, p=\frac{1}{1+Z_t\tau}\right)
\end{equation}
where $n$ and $p$ denote the size and probability of the NB distribution (R version 2.15.2), and $\tau$ denotes an over-dispersion parameter which will be estimated. (In the sense that the NB can be viewed as an over-dispersed Poisson process.)
The mean and variance of the NB distribution,
\begin{subequations}
\begin{eqnarray}
mean&=&\frac{n(1-p)}{p}=Z_t \\
variance&=&\frac{n(1-p)}{p^2}=Z_t(1+Z_t\tau)
\end{eqnarray}
\end{subequations}

When $\tau=0$, the NB distribution is reduced to a Poisson distribution.
Thus the likelihood for the week ($l_t$) can be simply calculated, namely the probability of observing $C_t$, given $Z_t$ and $\tau$, under the NB distribution \cite{Bret+09}. The overall likelihood function is
\begin{equation}
L(\theta|C_{0,...,N})=\prod\limits_{t=0}^N\,l_t
\end{equation}
where $\theta$ denotes the parameter vector. We used iterated filtering to estimate the maximum likelihood estimates for
$\theta$. This methodology has been extensively studied and used in a number of publications \cite{Ioni+06,Ioni+11,Earn+12,Cama+11,He+11,He+10,King+08}
\clearpage
\section{Influenza Vaccination Coverage in Different Countries}\label{s_vaccine}

\begin{table}[h!]
\begin{center}
\caption{Estimated influenza vaccination coverage in the general population}\vspace*{.1in}
\begin{tabular}{c|c|c|c|c|c}
\hline
Country/  &  &  & &\\
Population group & 2008-09  & 2009-10  & H1N1pdm (2009-10) & 2010-11 & 2011-12\\
\hline
Austria & & & 3 & &\\
Bulgaria & 4.74 & 6.33 & & 3.41 & 2.91\\
Cyprus & 11.8 & 16.1 & 3 & 11.9& 12.05\\
Czech Republic & 7.2 & 7.8  & 0.6 & &\\
Denmark & & & NA & &\\
England & & & NA & &\\
Estonia & & & 3 & 1.3 &\\
Finland & & & 50 & &\\
France &  & 20.6 & 8 & & \\
Germany & 28.1 & 26.6 & 8 & &\\
Greece & & & 3 & &\\
Hungary & & & 27 & 10.2 & 10.3\\
Iceland & 16.2 & 17.7 & 46 & 14.9 & 14.2\\
Ireland & & & 23 &  &\\
Italy & 19.1 & 19.6  & 4 & 17.2 & 17.8\\
Latvia & 0.9 & 0.8 & & 0.5 & 0.4\\
Lithuania  & 7.8 & 4 & & 3 & 6.4\\
Luxembourg & 16.6 &  18.1 & 6 & &\\
Malta & & & 23 &  &\\
The Netherlands & 21.9 & 22.4 & 30 & 21.3 & 65.7\\
Norway & 12 & 12(13) & 45 & 12 & 9\\
Poland & 4.1 & 3.1  & & 3.1 & 4.5\\
Portugal & 15 & 15(19) & 6 & 17.5 & 16.4\\
Romania & - & 5.2 & 9 & 5.6 & 3\\
Slovakia & 12.8 & 12.4 & 0.4 & 9.1 & 7.5\\
Slovenia & 7.3 & 7.3 & 5 & & 4.89\\
Spain & & & 27 & &\\
Sweden & & & 59 & &\\
\hline

\hline
\end{tabular}\\
\smallskip
Source:
\url{http://venice.cineca.org/Final_Seasonal_Influenza_Vaccination_Survey_2010.pdf}
\url{http://www.eurosurveillance.org/ViewArticle.aspx?ArticleId=20064}
\url{http://ecdc.europa.eu/en/escaide/materials/presentations\%202010/escaide2010_late_breakers_mereckiene.pdf}
\end{center}
\end{table}

\newpage

\begin{table}[h!]
\begin{center}
\caption{
Influenza vaccination coverage  in U.S. ($>$ 6 months) and Canada ($>$ 12 years old)} \label{T:vacc_US_CA}\vspace*{.1in}
\begin{tabular}{c|c|c|c|c|c}
\hline
Country & 2009-2010 (seasonal) & H1N1pdm & 2009-2010 (combined) & 2010-11 & 2011-12 \\
\hline
US & 41.2 & 27.2 & 47.8 & 43.0  & 41.8  (47.6?)  \\
Canada & 32.2 &	41.3 & NA	& 30.2	& 28.9 \\
\hline
\end{tabular}\\
\smallskip
Source:
\url{http://www.cdc.gov/flu/professionals/vaccination/coverage_0910estimates.htm.}
\url{http://www.statcan.gc.ca/pub/82-003-x/2010004/article/11348/tbl/tbl01-eng.htm},
\url{http://www.statcan.gc.ca/tables-tableaux/sum-som/l01/cst01/health101b-eng.htm}
\end{center}
\end{table}

In Japan, a study in four wards (communities) and one city in Tokyo showed that, among those between age 18 and 65,  38.1\% received
seasonal influenza vaccine and 12.1\% received A(H1N1pdm) influenza vaccine between October, 2009 and April, 2010 \cite{Yi+11}.
The vaccination coverage for H1N1pdm in Hong Kong is around 14\% \url{http://www.dh.gov.hk/}.

\clearpage
\section{Spatio-temporal Pattern from 1995 to 2005}\label{SI9505}

\begin{figure}[ht!]
\centerline{\includegraphics[width=12cm]{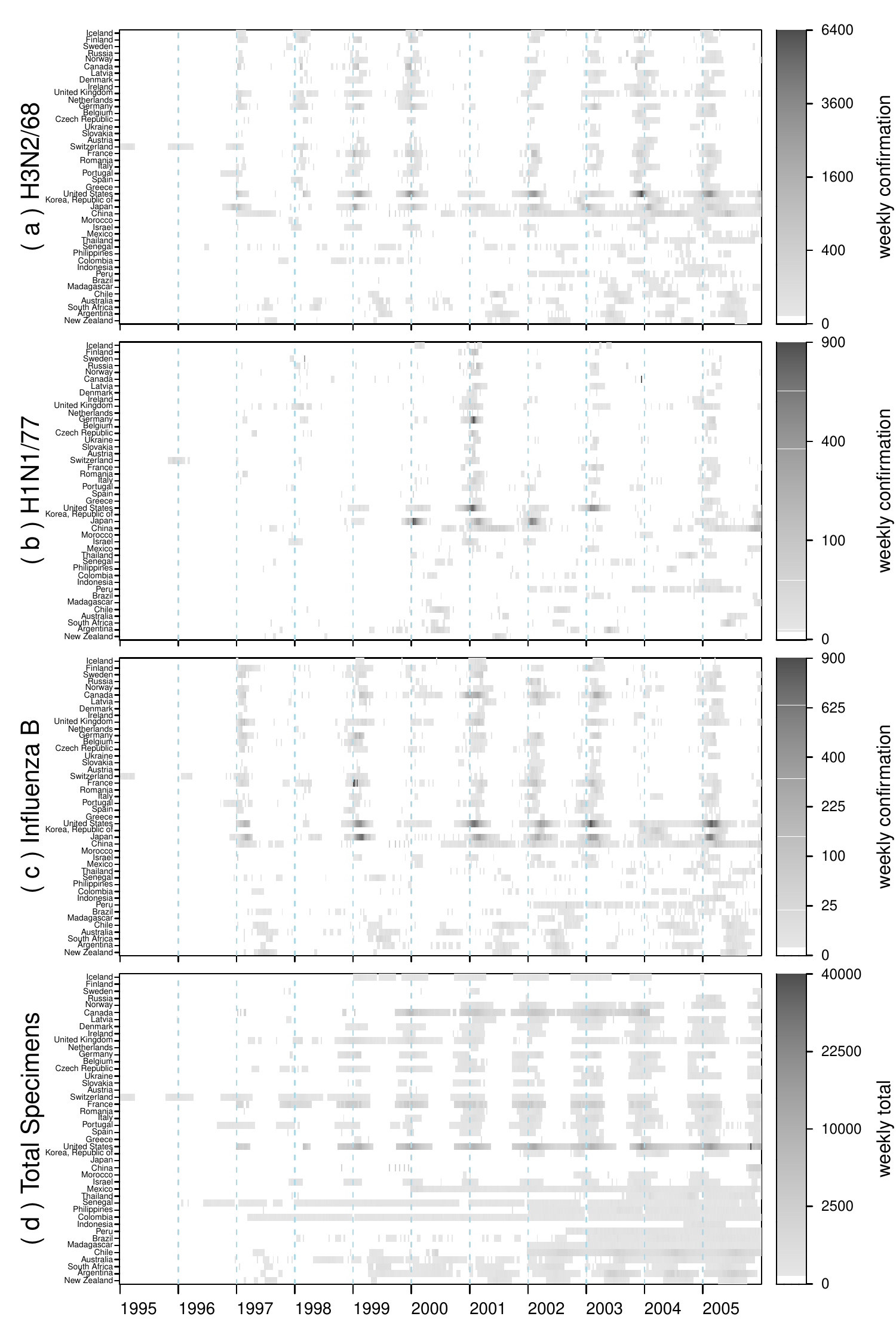}}
\caption{Spatio-temporal patterns of lab-confirmed cases of three types of influenza strains and total specimens processed for 44 countries which reported the largest confirmations between January, 1995 and December, 2005. Countries are listed in order of their latitudes. Grey scale shows the weekly lab-confirmed cases after taking square root. The horizontal axis is time in weeks.}
\label{Fig:spatialpre}
\end{figure}

Figure \ref{Fig:spatialpre} shows the spatio-temporal pattern of three strains from 1995 to 2005, and that of the weekly total specimens processed.

\section{Vaccine Components}\label{s_component}

Influenza virus vaccine composition is an indicator of predicted dominance of influenza strains.
Table \ref{vaccine} shows the influenza virus vaccine composition recommended by the Food and Drug Administration (FDA) of the United States (\url{ http://www.fda.gov/BiologicsBloodVaccines/GuidanceComplianceRegulatoryInformation/Post-MarketActivities/LotReleases/ucm062928.htm}),
which is usually the same as the recommendation by the World Health Organization for the north hemisphere countries (\url{http://www.who.int/influenza/vaccines/virus/recommendations/en/}).

\begin{table}[ht!]
\caption{Influenza vaccine composition recommended by U.S. Food and Drug Administration}\vspace*{.1in}\label{vaccine}
\begin{sideways}
{\tiny
\begin{tabular}{llll|lll}
\hline
\multicolumn{4}{c}{Vaccine Strains}&\multicolumn{3}{c}{UK Circulating Strains}\\
\hline
Season&A(H3N2)&A(H1N1)&B&A(H3N2)&A(H1N1)&B\\
\hline
1968/69&A/Hong Kong/1/68      &                    &                      &A/Hong Kong/1/68      &&\\
1969/70&A/Hong Kong/1/68      &                    &                      &A/Hong Kong/1/68      &&\\
1970/71&A/Hong Kong/1/68      &                    &                      &A/Hong Kong/1/68      &&\\
1971/72&A/Hong Kong/1/68      &                    &                      &A/Hong Kong/1/68      &&\\
       &A/England/42/72       &                    &                      &A/England/42/72       &                  &                          B/Hong Kong/5/72\\
1973/74&A/England/42/72       &                    &B/Victoria/98926/70   &A/Port Chalmers/1/73  &                  &                          B/Hong Kong/5/72\\
       &                      &                    &B/Hong Kong/5/72      &&&\\
1974/75&A/Port Chalmers/1/73  &                    &B/Hong Kong/5/72      &A/Port Chalmers/1/73  &                  &\\
1975/76&A/Port Chalmers/1/73  &A/Scotland/840/74   &B/Hong Kong/5/72      &A/Victoria/3/75       &                  &                          B/Hong Kong/5/72\\
1976/77&A/Victoria/3/75       &                    &B/Hong Kong/5/72      &A/Victoria/3/75       &                  &                          B/Hong Kong/8/73\\
1977/78&A/Victoria/3/75       &                    &B/Hong Kong/5/72      &A/Texas/1/77          &A/USSR/90/77      &\\
1978/79&A/Texas/1/77          &A/USSR/90/77        &B/Hong Kong/5/72      &A/Texas/1/77          &                  &                          B/Hong Kong/8/73\\
1979/80&A/Texas/1/77          &A/USSR/90/77        &B/Hong Kong/5/72      &A/Bangkok/1/79        &                  &                          B/Singapore/222/79\\
       &                      &                    &                      &                      &                  &                          B/Singapore/263/79\\
1980/81&A/Bangkok/1/79        &A/USSR/90/77        &B/Singapore/222/79    &A/Bangkok/1/79        &                  &                          B/Singapore/222/79\\
       &                      &A/Brazil/11/78      &&&&\\
1981/82&A/Bangkok/1/79        &A/Brazil/11/78      &B/Singapore/222/79    &A/Bangkok/1/79        &                  &\\
1982/83&A/Bangkok/1/79        &A/Brazil/11/78      &B/Singapore/222/79    &A/Philippines/2/82    &A/Brazil/11/78    &                          B/Singapore/222/79\\
1983/84&A/Philippines/2/82    &A/Brazil/11/78      &B/Singapore/222/79    &A/Philippines/2/82    &&\\
1984/85&A/Philippines/2/82    &A/Chile/1/83        &B/USSR/100/83         &A/Philippines/2/82    &&\\
1985/86&A/Philippines/2/82    &A/Chile/1/83        &B/USSR/100/83         &A/Mississippi/1/85    &&\\
1986/87&A/Christchurch/4/85   &A/Chile/1/83        &B/Ann Arbor/1/86      &A/Christchurch/4/85-  &&\\
       &A/Mississippi/1/85    &A/Singapore/6/86    &                      &                      &&\\
1987/88&A/Leningrad/360/86    &A/Singapore/6/86    &B/Ann Arbor/1/86      &A/Sichuan/2/87        &A/Singapore/6/86  &                          B/Beijing/1/87\\
1988/89&A/Sichuan/2/87        &A/Singapore/6/86    &B/Beijing/1/87        &A/Shanghai/11/87      &A/Singapore/6/86  &                          B/Beijing/1/87\\
1989/90&A/Shanghai/11/87      &A/Singapore/6/86    &B/Yamagata/16/88      &A/Shanghai/11/87      &A/Singapore/6/86  &\\
       &                      &                    &or B/Panama/45/90     &                      &                  &                          B/Yamagata/16/88\\
1990/91&A/Guizhou/54/89       &A/Singapore/6/86    &B/Yamagata/16/88      &A/Beijing/353/89      &A/Singapore/6/86  &                          B/Yamagata/16/88\\
       &                      &                    &or B/Panama/45/90     &                      &A/Victoria/36/88  &\\
1991/92&A/Beijing/353/89      &A/Singapore/6/86    &B/Yamagata/16/880     &A/Beijing/353/89      &A/Singapore/6/86  &                          B/Yamagata/16/88\\
       &                      &                    &or B/Panama/45/90     &                      &A/Victoria/36/88  &                          B/Panama/45/90\\
1992/93&A/Beijing/353/89      &A/Singapore/6/86    &B/Panama/45/90        &A/Beijing/32/92       &A/Singapore/6/86  &                          B/Panama/45/90\\
       &                      &                    &                      &                      &A/Victoria/36/88  &                          B/Quingdao/102/91\\
1993/94&A/Beijing/32/92       &A/Singapore/6/86    &B/Panama/45/90        &A/Shangdong/9/93      &                  &                          B/Panama/45/90\\
1994/95&A/Shangdong/9/93      &A/Singapore/6/86    &B/Panama/45/90        &A/Johnannesburg/33/94 &A/Singapore/6/86  &                          B/Beijing/184/93\\
       &                      &                    &                      &                      &A/Texas/36/91     &\\
1995/96&A/Johnannesburg/33/94 &A/Singapore/6/86    &B/Beijing/184/93      &A/Johnannesburg/33/94 &A/Singapore/6/86  &                          B/Beijing/184/93\\
       &                      &                    &                      &                      &A/Texas/36/91     &\\
1996/97&A/Wuhan/359/95        &A/Singapore/6/86    &B/Beijing/184/93      &A/Wuhan/359/95        &A/Bayern/7/95     &                          B/Beijing/184/93\\
1997/98&A/Wuhan/359/95        &A/Bayern/7/95       &B/Beijing/184/93      &A/Sydney/5/97         &A/Bayern/7/95     &                          B/Beijing/184/93\\
1998/99&A/Sydney/5/97         &A/Beijing/262/95    &B/Beijing/184/93      &A/Sydney/5/97         &A/Bayern/7/95     &                          B/Beijing/184/93\\
1999/00&A/Sydney/5/97         &A/Beijing/262/95    &B/Beijing/184/93      &A/Sydney/5/97         &A/Bayern/7/95     &                          B/Beijing/184/93\\
2000/01&A/Moscow/10/99        &A/New Caledonia/20/99&B/Beijing/184/93     &                      &A/New Caledonia/20/99&                       B/Sichuan/379/99\\
2001/02&A/Moscow/10/99        &A/New Caledonia/20/99&B/Sichuan/379/99     &A/Moscow/10/99        &A/New Caledonia/20/99 &   B/Sichuan/379/99 \\
       &                      &                     &                     &                      &    A/NewCaledonia/10/99 (H1N2) &\\
2002/03&A/Moscow/10/99        &A/New Caledonia/20/99&B/HongKong/330/2001  &A/Moscow/10/99        &A/New Caledonia/20/99&                       B/HongKong/330/2001\\
2003/04&A/Moscow/10/99        &A/New Caledonia/20/99&B/HongKong/330/2001  &A/Fujian/411/2002     &A/New Caledonia/20/99&\\
2004/05&A/Fujian/411/2002     &A/New Caledonia/20/99&B/Shanghai/361/2002  &A/Wellington/1/2004   &A/New Caledonia/20/99&                       B/Shanghai/361/2002\\
2005/06&A/California/7/2004   &A/New Caledonia/20/99&B/Shanghai/361/2002  &A/California/7/2004   &A/New Caledonia/20/99&                       B/HongKong/330/2001\\
2006/07&A/Wisconsin/67/2005   &A/New Caledonia/20/99&B/Malaysia/2506/2004 &A/Wisconsin/67/2005   &A/New Caledonia/20/99&\\
2007/08&A/Wisconsin/67/2005   &A/Solomon Islands/3/2006&B/Malaysia/2506/2004&A/Wisconsin/67/2005 &A/Solomon Islands/3/2006&                    B/Florida/4/2006\\
2008/09&A/Brisbane/10/2007    &A/Brisbane/59/2007  &B/Florida/4/2006      &A/Brisbane/10/2007    &A/Brisbane/59/2007&                          B/Florida/4/2006\\
       &                      &                    &                      &                      &A/California/07/2009 from Apr-2009&B/Malaysia/2506/2004\\
2009/10&A/Brisbane/10/2007    &A/Brisbane/59/2007  &B/Brisbane/60/2008    &A/Perth/16/2009       &A/California/07/2009&                        B/Brisbane/60/2008\\
2010/11&A/Perth/16/2009       &A/California/07/2009&B/Brisbane/60/2008    &A/Perth/16/2009       &A/California/07/2009&                        B/Brisbane/60/2008\\
2011/12&A/Perth/16/2009       &A/California/7/2009 &B/Brisbane/60/2008    &A/Perth/16/2009&A/California/7/2009& B-Victoria/B-Yamagata\\
2012/13&A/Victoria /361/2011  &A/California/7/2009 &B/Wisconsin /1/2010   &A/Victoria/361/2011&A/California/7/2009& B/Wisconsin/1/2010\\
2013/14&A/Texas/50/2012   &A/California/7/2009 &B/Massachusetts/2/2012 &A/Texas/50/2012&A/California/7/2009&    B-Yamagata\\
\hline
\end{tabular}
}
\end{sideways}
\end{table}
\url{http://www.hpa.org.uk/Topics/InfectiousDiseases/InfectionsAZ/SeasonalInfluenza/}

\clearpage

\section{Comparison among the US, UK and Canada}\label{s_usukcanada}

\begin{figure}[ht!]
\centerline{\includegraphics[width=12cm]{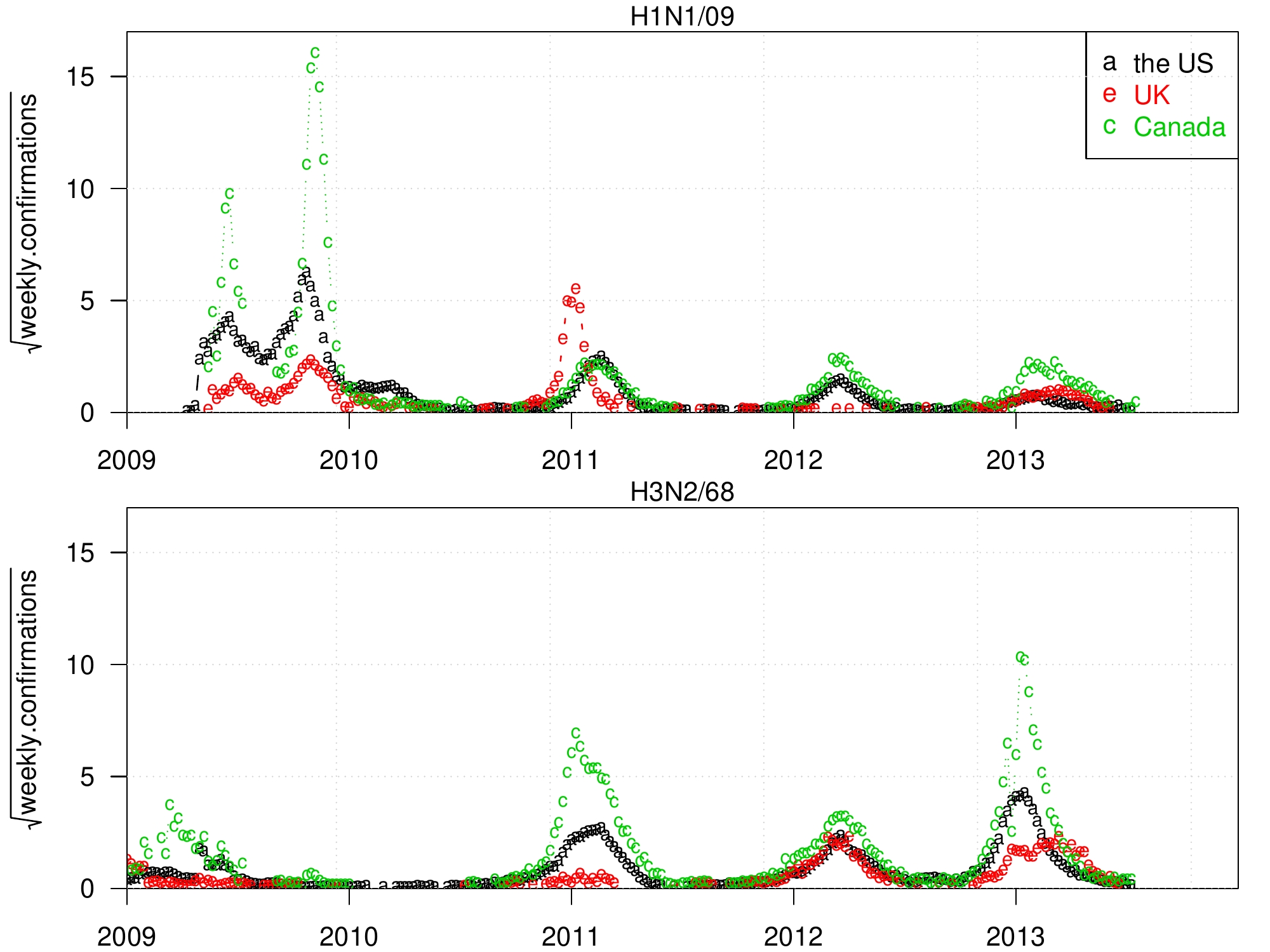}}
\caption{Comparison of the population standardized weekly confirmations between US, UK and Canada for H1N1pdm and H3N2.}
\label{comparison.US.UK.CA}
\end{figure}

Figure \ref{comparison.US.UK.CA} shows the comparison of  the standardized weekly confirmations among the United States, United Kingdom and Canada populations. These comparisons suggest that the attack rate in 2009 might be low in UK than in North America.

\clearpage
\section{A list of countries}\label{S:region}

Regions and sub-regions are from \url{http://millenniumindicators.un.org/unsd/methods/m49/m49regin.htm}.

\begin{table}[h!]
\caption{}
\begin{tabular}{rrrr}
\hline
name & subregion & pop2005 & latitude \\
\hline
Mauritius &  Eastern Africa & 1241173 & -20.255 \\
Madagascar &  Eastern Africa & 18642586 & -19.374 \\
Zambia &  Eastern Africa & 11478317 & -14.614 \\
Mozambique &  Eastern Africa & 20532675 & -14.422 \\
United Republic of Tanzania &  Eastern Africa & 38477873 & -6.270 \\
Rwanda &  Eastern Africa & 9233793 & -1.998 \\
Kenya &  Eastern Africa & 35598952 &  0.530 \\
Uganda &  Eastern Africa & 28947181 &  1.280 \\
Ethiopia &  Eastern Africa & 78985857 &  8.626 \\
Angola &  Middle Africa & 16095214 & -12.296 \\
Democratic Republic of the Congo &  Middle Africa & 58740547 & -2.876 \\
Congo &  Middle Africa & 3609851 & -0.055 \\
Cameroon &  Middle Africa & 17795149 &  5.133 \\
Central African Republic &  Middle Africa & 4191429 &  6.571 \\
Chad &  Middle Africa & 10145609 & 15.361 \\
Sudan &  Northern Africa & 36899747 & 13.832 \\
Egypt &  Northern Africa & 72849793 & 26.494 \\
Algeria &  Northern Africa & 32854159 & 28.163 \\
Morocco &  Northern Africa & 30494991 & 32.706 \\
Tunisia &  Northern Africa & 10104685 & 35.383 \\
South Africa &  Southern Africa & 47938663 & -30.558 \\
Cote d'Ivoire &  Western Africa & 18584701 &  7.632 \\
Ghana &  Western Africa & 2253501 &  7.960 \\
Sierra Leone &  Western Africa & 5586403 &  8.560 \\
Togo &  Western Africa & 6238572 &  8.799 \\
Nigeria &  Western Africa & 141356083 &  9.594 \\
Guinea &  Western Africa & 9002656 & 10.439 \\
Guinea-Bissau &  Western Africa & 1596929 & 12.125 \\
Burkina Faso &  Western Africa & 13933363 & 12.278 \\
Senegal &  Western Africa & 1177034 & 15.013 \\
Cape Verde &  Western Africa & 506807 & 15.071 \\
Mali &  Western Africa & 1161109 & 17.350 \\
\hline
\end{tabular}

\end{table}

\begin{table}[h!]
\caption{}
\begin{tabular}{rrrr}
\hline
name & subregion & pop2005 & latitude \\
\hline
Niger &  Western Africa & 1326419 & 17.426 \\
Mauritania &  Western Africa & 2963105 & 20.260 \\
United States &  Northern America & 299846449 & 39.622 \\
Canada &  Northern America & 32270507 & 59.081 \\
Saint Lucia &  Caribbean & 16124 & 13.898 \\
Martinique &  Caribbean & 395896 & 14.653 \\
Guadeloupe &  Caribbean & 438403 & 16.286 \\
Saint Martin &  Caribbean & 0 & 18.094 \\
Jamaica &  Caribbean & 2682469 & 18.151 \\
Dominican Republic &  Caribbean & 9469601 & 19.015 \\
Cuba &  Caribbean & 11259905 & 21.297 \\
Panama &  Central America & 3231502 &  8.384 \\
Costa Rica &  Central America & 4327228 &  9.971 \\
Nicaragua &  Central America & 5462539 & 12.840 \\
El Salvador &  Central America & 6668356 & 13.736 \\
Honduras &  Central America & 683411 & 14.819 \\
Guatemala &  Central America & 12709564 & 15.256 \\
Mexico &  Central America & 104266392 & 23.951 \\
Argentina &  South America & 38747148 & -35.377 \\
Uruguay &  South America & 3325727 & -32.800 \\
Chile &  South America & 16295102 & -23.389 \\
Paraguay &  South America & 5904342 & -23.236 \\
Bolivia &  South America & 9182015 & -16.715 \\
Brazil &  South America & 186830759 & -10.772 \\
Peru &  South America & 27274266 & -9.326 \\
Ecuador &  South America & 13060993 & -1.385 \\
Colombia &  South America & 4494579 &  3.900 \\
French Guiana &  South America & 192099 &  3.924 \\
Suriname &  South America & 452468 &  4.127 \\
Venezuela &  South America & 26725573 &  7.125 \\
Kyrgyzstan &  Central Asia & 5203547 & 41.465 \\
Uzbekistan &  Central Asia & 26593123 & 41.750 \\
\hline
\end{tabular}

\end{table}

\begin{table}[h!]
\caption{}
\begin{tabular}{rrrr}
\hline
name & subregion & pop2005 & latitude \\
\hline
Kazakhstan &  Central Asia & 15210609 & 48.160 \\
China &  Eastern Asia & 1312978855 & 33.420 \\
Japan &  Eastern Asia & 127896740 & 36.491 \\
Korea, Republic of &  Eastern Asia & 47869837 & 36.504 \\
Mongolia &  Eastern Asia & 2580704 & 46.056 \\
Sri Lanka &  Southern Asia & 19120763 &  7.612 \\
India &  Southern Asia & 1134403141 & 21.000 \\
Bangladesh &  Southern Asia & 15328112 & 24.218 \\
Bhutan &  Southern Asia & 637013 & 27.415 \\
Nepal &  Southern Asia & 27093656 & 28.253 \\
Pakistan &  Southern Asia & 158080591 & 29.967 \\
Iran (Islamic Republic of) &  Southern Asia & 69420607 & 32.565 \\
Afghanistan &  Southern Asia & 25067407 & 33.677 \\
Indonesia &  South-Eastern Asia & 226063044 & -0.976 \\
Singapore &  South-Eastern Asia & 4327468 &  1.351 \\
Malaysia &  South-Eastern Asia & 25652985 &  4.201 \\
Philippines &  South-Eastern Asia & 84566163 & 11.118 \\
Cambodia &  South-Eastern Asia & 13955507 & 12.714 \\
Thailand &  South-Eastern Asia & 63002911 & 15.700 \\
Lao People's Democratic Republic &  South-Eastern Asia & 566391 & 19.905 \\
Viet Nam &  South-Eastern Asia & 85028643 & 21.491 \\
Oman &  Western Asia & 2507042 & 21.656 \\
Qatar &  Western Asia & 796186 & 25.316 \\
Bahrain &  Western Asia & 724788 & 26.019 \\
Jordan &  Western Asia & 5544066 & 30.703 \\
Israel &  Western Asia & 6692037 & 31.026 \\
Iraq &  Western Asia & 27995984 & 33.048 \\
Syrian Arab Republic &  Western Asia & 18893881 & 35.013 \\
Turkey &  Western Asia & 72969723 & 39.061 \\
Azerbaijan &  Western Asia & 8352021 & 40.430 \\
Armenia &  Western Asia & 3017661 & 40.534 \\
Georgia &  Western Asia & 4473409 & 42.176 \\
\hline
\end{tabular}

\end{table}

\begin{table}[h!]
\caption{}
\begin{tabular}{rrrr}
\hline
name & subregion & pop2005 & latitude \\
\hline
Bulgaria &  Eastern Europe & 7744591 & 42.761 \\
Romania &  Eastern Europe & 21627557 & 45.844 \\
Hungary &  Eastern Europe & 10086387 & 47.070 \\
Republic of Moldova &  Eastern Europe & 3876661 & 47.193 \\
Slovakia &  Eastern Europe & 5386995 & 48.707 \\
Ukraine &  Eastern Europe & 46917544 & 49.016 \\
Czech Republic &  Eastern Europe & 10191762 & 49.743 \\
Poland &  Eastern Europe & 38195558 & 52.125 \\
Belarus &  Eastern Europe & 9795287 & 53.540 \\
Russia &  Eastern Europe & 143953092 & 61.988 \\
United Kingdom &  Northern Europe & 60244834 & 53.000 \\
Ireland &  Northern Europe & 4143294 & 53.177 \\
Lithuania &  Northern Europe & 3425077 & 55.336 \\
Denmark &  Northern Europe & 5416945 & 56.058 \\
Latvia &  Northern Europe & 2301793 & 56.858 \\
Estonia &  Northern Europe & 1344312 & 58.674 \\
Norway &  Northern Europe & 4638836 & 61.152 \\
Sweden &  Northern Europe & 9038049 & 62.011 \\
Finland &  Northern Europe & 5246004 & 64.504 \\
Iceland &  Northern Europe & 295732 & 64.764 \\
Malta &  Southern Europe & 402617 & 35.890 \\
Greece &  Southern Europe & 11099737 & 39.666 \\
Spain &  Southern Europe & 43397491 & 40.227 \\
Portugal &  Southern Europe & 10528226 & 40.309 \\
Albania &  Southern Europe & 3153731 & 41.143 \\
The former Yugoslav Republic of Macedonia &  Southern Europe & 2033655 & 41.600 \\
Italy &  Southern Europe & 5864636 & 42.700 \\
Serbia &  Southern Europe & 9863026 & 44.032 \\
Bosnia and Herzegovina &  Southern Europe & 3915238 & 44.169 \\
Croatia &  Southern Europe & 455149 & 45.723 \\
Slovenia &  Southern Europe & 1999425 & 46.124 \\
France &  Western Europe & 60990544 & 46.565 \\
\hline
\end{tabular}

\end{table}

\begin{table}[h!]
\caption{}
\begin{tabular}{rrrr}
\hline
name & subregion & pop2005 & latitude \\
\hline
Switzerland &  Western Europe & 7424389 & 46.861 \\
Austria &  Western Europe & 8291979 & 47.683 \\
Luxembourg &  Western Europe & 456613 & 49.771 \\
Belgium &  Western Europe & 10398049 & 50.643 \\
Germany &  Western Europe & 82652369 & 51.110 \\
Netherlands &  Western Europe & 1632769 & 52.077 \\
New Zealand &  Australia and New Zealand & 4097112 & -42.634 \\
Australia &  Australia and New Zealand & 20310208 & -24.973 \\
New Caledonia &  Melanesia & 234185 & -21.359 \\
Fiji &  Melanesia & 828046 & -17.819 \\
\hline
\end{tabular}

\end{table}

\end{document}